 \documentclass{aa}

\begin{document}

 \thesaurus{11 (11.19.2; 11.19.6; 11.11.1; 11.09.1 \object{NGC
 922}; 03.13.4)}

 \title{Dust-penetrated morphology in the high-redshift universe: clues
 from \object{NGC 922} }

 \author{David L. Block \inst{1},
         Iv\^anio Puerari \inst{2},
         Marianne Takamiya \inst{3},
         Roberto Abraham \inst{4},
         Alan Stockton \inst{5},
         Ian Robson \inst{6}
         \and
         Wayne Holland \inst{6}}

 \institute{Dept of Computational and Applied Mathematics,
            University of the Witwatersrand,
            Private Bag 3, WITS 2050, South Africa
           \and
            Instituto Nacional de Astrof\'\i sica,
            Optica y Electr\'onica,
            Calle Luis Enrique Erro 1, 72840 Tonantzintla, Puebla, M\'exico
            \and
            Gemini Observatory, 670 North Aohoku Place,
            Hilo, HI 96720, USA
            \and
            Department of Astronomy, University of Toronto,
            60 St. George Str., Toronto,  ON M5S 3H8, Canada
            \and
            Institute for Astronomy, University of Hawaii, 2680 Woodlawn
            Drive, Honolulu, HI 96822, USA
            \and
            Joint Astronomy Centre, Univ. Park, 660 North Aohoku Place,
            Hilo, HI 96720, USA}

 \offprints{D.L. Block}

 \date{Received .................. / Accepted ..................}

 \titlerunning{\object{NGC 922}: A Rosetta Stone?}
 \authorrunning{D.L. Block et al.}

 \maketitle

 \begin{abstract}

 Results from the Hubble Deep Field (HDF) North and South show a
 large percentage of high-redshift galaxies whose appearance falls
 outside traditional classification systems. The nature of these
 objects is poorly understood, but sub-mm observations indicate
 that at least some of these systems are heavily obscured (Sanders
 \cite{sanders00}). This raises the intriguing possibility that a
 physically meaningful classification system for high-redshift
 galaxies might be more easily devised at rest-frame infrared
 wavelengths, rather than in the optical regime. Practical
 realization of this idea will become possible with the advent of
 the {\em Next Generation Space Telescope} (NGST). In order to
 explore the capability of NGST for undertaking such science, we
 present NASA-IRTF and SCUBA observations of \object{NGC 922}, a
 chaotic system in our local Universe which bears a striking
 resemblance to objects such as HDF 2-86 ($z=0.749$) in the HDF
 North.  If objects such as \object{NGC 922} are common at
 high-redshifts, then this galaxy may serve as a local
 morphological `Rosetta stone' bridging low and high-redshift
 populations. In this paper we demonstrate that quantitative
 measures of galactic structure are recoverable in the rest-frame
 infrared for \object{NGC 922} seen at high redshifts using NGST,
 by simulating the appearance of this galaxy at redshifts z=0.7 and
 z=1.2 in rest-frame K$'$. While this object cannot be classified
 within any optical Hubble bin, simulated NGST images at these
 redshifts can be readily classified using the dust penetrated z
 $\sim$ 0 template of Block and Puerari (\cite{blockpuerari99}) and Buta
and Block (\cite{butablock01}). The near-infrared disk of NGC 922
is not peculiar at all; rather, it is remarkably {\em regular},
even presenting spiral
arm modulation, a characteristic signature of several grand design
galaxies.
 Our results suggest that the capability of {\em efficiently}
 exploring
 the rest-wavelength IR morphology of high-z galaxies should probably be
 a key factor in deciding the final choice of instruments for the NGST.


 \end{abstract}

 \section{Introduction}

 Considerable progress has recently been made in establishing the
 morphological mix of field galaxy populations as a function of
 redshift  (eg. Glazebrook et al. \cite{glazebrook95}, Driver et al.
 \cite{driver98}, Abraham et al. \cite{abraham96a}). As a by-product of
 this work, a number
 of investigations probing the systematic changes in galaxy morphology
 as a function of rest-wavelength have been undertaken
(Abraham et al. \cite{abraham96b}, Bouwens et al. \cite{bouwensetal98},
 Giavalisco et al. \cite{giavaliscoetal96}).
 These studies have focused primarily on
 changes
 which occur as the rest-wavelength of observation moves from the optical
 regime into the near-ultraviolet (so-called ``morphological
 K-corrections''). Remarkably little quantitative work has explored
 the systematic variations in morphology which occur as one moves
 from the optical in the {\em other} wavelength direction, namely
 into the near-IR. This is rather surprising, since the nearly
 constant mass-to-light ratio as a function of age for simple
 stellar populations suggests that near-IR morphology probes
 underlying stellar mass much more fairly than does optical
 morphology  (Kauffmann and Charlot \cite{kauffmann98}; fig. 4
 in Charlot \cite{charlot98};  Charlot \cite{charlot96}; Frogel et al.
 \cite{frogel96}).

 The morphology of local galaxies can completely change once
 Population I disks are dust penetrated. One of many examples is
 \object{ NGC 5195}, an optical irregular which presents an SBa
 morphology in the near-infrared. The extinction at K$'$ (2.1$\mu
 m$) is  only 10\% that in the V (0.55$\mu m$) band. For local
 field galaxies, there is no correlation between dust penetrated
 classes and optical Hubble types; the Hubble tuning fork does not
 constrain the morphology of the old stellar Population II disks
 (Block and Puerari \cite{blockpuerari99}). Less is known at higher
 redshifts, since the evolved stellar disks of high-z galaxies in
 the HDF have never been explored at restframe K$'$ band.  Images
 of galaxies with redshifts ${\rm z}\sim 0.5 - 1$ or higher secured
 using the Hubble Space Telescope and NICMOS never penetrate the
 dusty, gaseous Population I mask. At $z>3$, even H band (1.6$\mu
 m$) observed flux stems from emission shortward  of 4000 \AA.
 Early results from SCUBA suggest that a substantial proportion of
 high-redshift star-formation is occurring in heavily obscured
 galaxies (Sanders \cite{sanders00}; see also fig. 5 in Hughes
 \cite{hughes96}).

 Fortunately, upcoming advances in technology will soon allow
 the rest-frame IR morphology of galaxies to be probed at both low
 and high-redshifts. Near-infrared imaging surveys of local
 galaxies (such as the IR component of the VISTA survey) will be
 entirely comparable to those currently achievable at optical
 wavelengths. Similarly, the {\em Next Generation Space Telescope}
 (NGST) will allow studies of rest-wavelength IR morphology to be
 extended to high-redshifts. In this regard, the synergy between
 VISTA and NGST may turn out to be rather similar to that currently
 existing between ground-based optical facilities and HST.

 Motivated by this possibility, we have begun an investigation of
 the capability of NGST for extending local morphological studies
 to high redshifts. In this preliminary investigation, our strategy
 will be to focus on a single extreme case (NGC 922), which
 highlights the potential benefits of extending quantitative
 analyses of rest-wavelength near-IR morphology to high redshifts.
 This object is particularly well-suited for our purposes, since
 (in common with numerous high-redshift galaxies) ``The morphology
 of \object{NGC 922} is so peculiar as to be outside the
 classification system. It would be called a sport by
 nineteenth-century animal breeders'' (see Sandage and Bedke
 \cite{sandagebedke94}, Panel 313). This galaxy bears a striking
 optical resemblance to chaotic objects seen at high redshifts such
 as HDF 2-86 (van den Bergh \cite{vandenbergh98}) at $z=0.749$ in
 the Hubble Deep Field (HDF), which van den Bergh et al.
 (\cite{vandenbergh96}) suggest might be a spiral in the process of
 being assembled. Both \object{NGC 922} and HDF 2-86 contain a
 prominent arc-like feature in the one-half of the galaxy in which
 star formation is apparently proceeding at a vigorous rate (see
 Table 1 in Cohen \cite{cohen76} and Devereux \cite{devereux89}).


 %


 \section{Simulations of \object{NGC 922} with NGST at Restframe
 K$'$}

 $K'$ images of \object{NGC 922} were obtained at the NASA Infrared
 Telescope Facility with NSFCam, operating in the 0.3 arcsec mode,
 giving a field of 76\arcsec.  NGC\,922 was placed successively in
 each quadrant of the detector in order to have good coverage of
 nearly uncontaminated blank sky, from which a median averaged sky
 flat could be obtained.  Small dithers superposed on this basic
 4-point pattern ensured that stars in the field fell at different
 points of the detector each time.  In all, 17 integrations of 1
 minute each were obtained. Our K$'$ mosaic appears in fig. 1.

 The galaxy was also observed at 850$\mu m$ using the Submillimetre
 Common-User Bolometer Array (SCUBA; see Holland et al. \cite{holland99})
 at the James Clerk Maxwell Telescope (JCMT) at Mauna Kea.
 The observing mode used was jiggle-mapping in which the
 secondary mirror of the telescope compensates for the
 instantaneous under-sampled images by offsetting the array
 position to produce a Nyquist-sampled image. Some
 300 separate `jiggle-maps' (or integrations)  were
 secured over a three night period, with a total on-source integration
 time of 2.7 hours.
 The data was reduced using the SURF software package (Jenness and
Lightfoot \cite{jenness98}),
 correcting for noisy pixels, flatfield variations, atmospheric opacity,
 and short-time sky variations. The data were flux calibrated using
 beam-maps of Uranus and CRL618. The peak flux in the map is
 25.0$\pm$1.7 mJy/beam (following smoothing by a 7$''$ gaussian)
 and the integrated flux over a 100$''$ diameter aperture, centered on
 the peak, is estimated to be $\sim$ 36mJy. Absolute calibration accuracy
 is estimated to be about 6\%.

 Emission at 850$\mu m$ is not
 confined to only one-half of the disk in which star-formation is
 occurring. The dust morphology is {\em radically
 different}
 to the near-infrared NASA-IRTF morphology (see fig. 2); cold dust
grains, seen in
 emission at 850$\mu m$, are
 not restricted to the bright one-half of the almost
 'sliced-in-two' disk.

 Assuming a mass absorption
 coefficient of 0.007 at 850$\mu m$ and a grain temperature of 10-20K
 (consistent with cold dust grain temperatures
 inferred using radiative transfer codes e.g. Block et al.
 \cite{blocketal94a}), we compute an atomic and molecular hydrogen
 gas mass of $\sim$ 2-4 10$^{9}$ M$_{\odot}$. The dust
 mass for \object{NGC 922} of $\sim$ 10$^{7}$ M$_{\odot}$, one order of
 magnitude greater than that of the `Evil Eye' Galaxy \object{NGC
 4826} (often cited as a prototype of a very dusty galaxy).


 Our artificially redshifted images of the dusty \object{NGC 922} are
 based on the preliminary description of the NGST facility given in
 Gillett and Mountain (\cite{gillettmountain98}). For concreteness,
 we assumed an 8m telescope. Our images assume
 an imager corresponding to the design of an 8192 $\times$ 8192
 pixel format near-infrared camera operating on NGST from 0.6 to
 5.3$\mu m$ (which includes the L and M bands), as described in
 Bally and Morse (\cite{ballymorse99}). It is of course emphasized
 that at the present time the NGST telescope and camera designs
 have not been finalized, and the orbit placement of NGST not yet
 been decided. Indeed the present investigation of what can be
 achieved with the currently planned facility may provide some
 useful guidance to instrument designers.

 In order to probe galaxies using NGST at redshifts 0.7 and 1.2 in
 their K$'$ restframes, imaging detectors in the broadband L
 (3.7$\mu m$; bandpass $\delta$L=0.65$\mu m$) and M (4.7$\mu m$;
 $\delta$M=0.45$\mu m$) regimes are required.  At M, for example,
 the sky background at an excellent groundbased site such as Mauna
 Kea is 10 million times brighter than it will be for NGST (see
 fig. 1 in Gillett and Mountain \cite{gillettmountain98}, where a
 sky background at Mauna Kea of $\sim$ 305 Jansky arcsec$^{-2}$ is
 adopted, but only 1.9$\times$10$^{-5}$ Jansky arcsec$^{-2}$ for
 NGST).

 We follow the simulation methodology of Takamiya
 (\cite{takamiya99}).  The simulations recreate the images of
 \object{NGC 922} when moved out to higher redshifts, always in a
 pre-selected restframe. In this paper, the images of
 \object{NGC 922} are simulated at redshifts z=0.7 and z=1.2 in the
 dust penetrated K$'$ rest-frame. Since the rest-frames are
 matched, no pixel $k$ corrections are applied (Takamiya
 \cite{takamiya99}). If this was not the case, then $k$ corrections
 in simulations are necessary (e.g., Bouwens et al.
 \cite{bouwensetal98}, Giavalisco et al. \cite{giavaliscoetal96}).
 Furthermore, no spectral energy distribution templates are needed:
 these are only required for rest-frames which are not matched,
 such as when simulating HDF rest-frame UV images from local
 \hbox{z $\sim$ 0} optical ones (see for example, Abraham et al.
 \cite{abraham96b}).




 We assume that the NGST sky background is about
 twice the minimum background observed by COBE (Hauser
 \cite{hauser94}), corresponding to a sky surface brightness at L
 and M of 19.55 and 17.31 magnitudes per square arcsecond,
 respectively. (From good near-infrared groundbased sites, the
 corresponding sky surface backgrounds at L and M are \hbox{$\sim$
 5.99} mag arcsec$^{-2}$ and --0.24 mag arcsec$^{-2}$,
 respectively.) We furthermore assume pixel sizes of 50
 milliarcseconds, a gain of 4 electrons per ADU (analog digital
 unit), a readout noise of 4 electrons and an output point spread
 function of gaussian distribution, with a full width at half
 maximum (FWHM) of 0$\farcs$12. Assuming eight optical surfaces
 with 95\% transmission (these include the primary and secondary
 mirrors as well as lenses and filters in the detector) and a 40\%
 detector quantum efficiency, the system throughput is taken to be
 26\%. An on-source integration time of one hour was adopted.
 Results from these simulations are shown in the left hand column
 of fig. 3. A morphological decomposition based on these images
 is shown in the right hand column of this figure, and will be
 described in the next section.  As a comparison, we ran a set of
 simulations on a galaxy which {\it can} readily be classified in the
 optical regime: \object{NGC 2997}. NGC 2997 shows a
 magnificent
 grand design structure in the optical regime; it is of Hubble type Sc
 and belongs to van den Bergh luminosity class I bin. The results of
 the NGC 2997 simulations are presented in fig. 4.


 \section{Morphological classification}

 In the present paper we adopt the dust penetrated classification
 scheme for local spirals proposed by Block and Puerari
 (\cite{blockpuerari99}) and Buta and Block (\cite{butablock01}).
 In the near-infrared, the morphology of the older star dominated
 disk indicates a simple classification scheme (see figs. 5 and 7):
H$m$
 (where $m$ is the dominant Fourier harmonic), three pitch angle
 classes $\alpha$, $\beta$ and $\gamma$ and thirdly, a bar strength
 parameter. Bar strengths are derived from the gravitational
 potential or torque of the bar, and not from bar ellipticity. A
 bar which might be highly elongated (and therefore `strong' in
 optical images) may sometimes have a negligible bar strength, as
 inferred from its mass distribution and potential in near-infrared
 images (Buta and Block \cite{butablock01}). We will show that
 the dust-penetrated tuning fork template for local galaxies shown
 in fig. 7 readily accommodates \object{NGC 922}.

 The 2-D Fast Fourier decomposition of \object{NGC 922} images uses
 a program developed by I. Puerari (Schr\"oder et al.
 \cite{schroder94}). Following Danver (\cite{danver42}) and Kalnajs
 (\cite{kalnajs75}), logarithmic spirals of the form $r=r_o {\rm
 exp} (-m \theta /p_{\rm max})$ are employed in the decomposition.

 The amplitude of each Fourier component is given by (Schr\"oder et
 al. \cite{schroder94})

 $$\displaystyle A(m,p) = \frac{\Sigma_{i=1}^I \Sigma_{j=1}^J
 I_{ij} ({\rm ln}\;r,\theta)\; {\rm exp}\;(-i(m \theta + p\; {\rm
 ln}\;r))}{\Sigma_{i=1}^I \Sigma_{j=1}^J I_{ij} ({\rm
 ln}\;r,\theta)}$$

 \noindent where $r$ and $\theta$ are polar coordinates, $I({\rm
 ln}\;r,\theta)$ is the intensity at position $({\rm ln}\;r,
 \theta)$, $m$ represents the number of arms or modes, and $p$ is
 the variable associated with the pitch angle $P$, defined by
 $\displaystyle \tan P = -\frac{m}{p_{\rm max}}$.

 After having deprojected the K$'$ images and identifying the
 dominant modes, we calculate the inverse Fourier transform, as
 follows:

 We define the variable $\displaystyle u \equiv {\rm ln}\;r$. Then

 $$\displaystyle S(u,\theta) = \sum_m S_m (u) {\rm e}^{im \theta}
 $$

 where

 $$\displaystyle S_m(u) = \frac{D}{{\rm e}^{2u} 4 \pi^2}
 \int_{-\infty}^ {+\infty}\;G_m (p) A(p,m) {\rm e}^{i p u}\;dp$$

 and

 $$\displaystyle D=\Sigma_{i=1}^I \Sigma_{j=1}^J I_{ij} (u,
 \theta)$$

 $G_m(p)$ is a high frequency filter used by Puerari and Dottori
 (\cite{pueraridottori92}). For the spiral with $\displaystyle \tan
 P = -\frac{m}{p_{max}^m}$ it has the form

 $$\displaystyle G_m(p) = {\rm exp} \left[ -\frac{1}{2} \left(
 \frac{p - p_{max}^m}{25} \right)^2 \right] $$

 \noindent where $p_{\rm max}^m$ is the value of $p$ for which the
 amplitude of the Fourier coefficients for a given $m$ is maximum.
 This filter is also used to smooth the $A(p,m)$ spectra at the
 interval ends (see Puerari and Dottori \cite{pueraridottori92}).

\object{NGC 922} is close to face-on; deep optical CCD images
present
 an almost circular arc/disk. No deprojection corrections were
 therefore necessitated. Fourier
 spectra were determined for \object{NGC 922} at
 z=0.7 and z=1.2 (see fig. 8),  and inverse Fourier transform
contours were
 then overlayed on the simulated restframe K$'$ mosaics, following
 the prescription above.
 The right hand columns of figs. 3 and 4 show
 the results of the inverse Fourier contour overlays on our one-hour
NGST simulations.

 The Fourier spectra of NGC 922 (see fig. 8) shows a
 strong lopsided m=1 component, with a pitch angle
 corresponding to 39$^{\circ}$ (ie. class $\gamma$ in fig. 5).
The m=1 component is
 associated with the arc so clearly seen in our NASA-IRTF
 image.
 The m=2 component is almost as strong in
 amplitude as the m=1 component (fig. 8), and corresponds to the
 bisymmetric spiral seen at K$'$ which lies within the arc (fig. 1).
What is
 intriguing about the m=2 component in fig. 8 is that it is doubly
peaked at
 opposite sides of the origin of the $p$-axis, indicative of a set of
 leading
 and trailing wave modes which are propagating inward and outward, being
 reflected off a central bulge (see Puerari et al. \cite{puerarietal00}).
 {\it In other words, \object{NGC 922} shows
 the unmistakable signature of spiral arm modulation in its stellar
 disk}. Spiral arm modulation
 has hitherto only been detected in grand design disks (such as Messier
 81). With these beating mode considerations, \object{NGC 922}
 could well be a grand design spiral in the process of assembly.
 In our template of figs. 5 and 7, the bisymmetric spiral in
 \object{NGC 922} would carry the classification H2$\gamma$.

 \object{NGC 922} has a gravitational bar strength or torque
 (Buta and Block \cite{butablock01}) of 2 (on a scale of 0 to 6).
 The
 complete quantitative classification of \object{NGC 922} (focussing on
 the inner spiral seen at K$'$) is therefore H2$\gamma$2.

 The bar
 strength component
 of this classification scheme is particularly interesting, given the
 evidence for a rapid decline in the proportion of {\em optically}
 barred galaxies at high redshifts (Abraham et al. \cite{abraham99};
 van den Bergh et al. \cite{sidney00}). The corresponding behaviour
 in the
 rest-frame infrared, where bars are much more prominent in local data
 (Eskridge et al. \cite{eskridge00}), is unknown.

 It must be stressed that Fourier spectra of \object{NGC 922} are
 being generated on square ``postage stamp'' FITS images only 3$''$
 on a side (64$\times$64 pixels, where 1 pixel corresponds to 50
 mas), yet the appearance of the modes are strikingly similar to
 those of the local $z\sim 0$ spectra. The decrease of spatial
 resolution in higher redshift space does not wash out the dominant
 modes of \object{NGC 922} seen locally (fig. 8).
 In fact, given two sets of Fourier spectra, one for the galaxy at
 its original distance, and another, generated from the simulated
 image at z=0.7 or z=1.2, it is a challenge to find any noticeable
 difference in the dominant modes at all (independent of q$_0$).
 Identical conclusions pertain to the preservation of pitch angle in NGC
2997 (see fig. 9). The class $\beta$ bin is retained throughout.

An interesting point to note is that is that the first and second
order harmonics
in figs. 8 and 9 are not mutually exclusive. A galaxy
often presents both H1 and H2 components. Modal theories of spiral
structure
predict that m=1 modes should generally be accompanied by m=2 modes when
available, since the latter are more efficient in transporting angular
momentum outwards. For example, the inner leading arm in NGC 4622
(labeled in fig. 6a of Block et al. \cite{blocketal94b})
co-exists with an outer set of m=2 spirals. Block et al.
(\cite{blocketal00}) classify the lopsided arm in NGC 4622 as
H1$\alpha$ and the outer set of tightly wrapped spirals as H2$\alpha$.
What fig. 8 and 9 do show is part of a general trend already
noted
by Block et al. (\cite{blocketal94b}) that modes higher than 2 (ie.
multi-arm
features)  are generally suppressed in stellar disks imaged at
restframe K$'$, whereas the
dusty, gaseous components seen in optical images are dynamically very
active,
fueling Jeans instability and readily supporting multi-arm features.


 A note on deprojections at high-z is in order. By the time NGST is
 launched, IFU-based spectroscopy ought to be the norm (in fact, some
 form of resolved spectroscopic capability is likely to be a component of
 the NGST instrument load). Kinematical information can then be
 used,
 as in our local Universe, to secure accurate deprojection parameters.
 However, we do wish to point out that
 lessons from z$\sim$0 galaxies show that deprojections
 based on axial ratios alone are excellent (kinematical studies
 simply finely-tune them). In our high-z universe, axial ratios
 from even relatively low signal-noise NGST images could readily be
determined
 from image moments; being based on calculations summed over
 many -- rather than a few -- pixels,
 they should be
 remarkably robust (see Frei \cite{frei00}), even without
 spectroscopic/kinematical NGST information.

 When high redshift galaxies are observed at the L and M bands with
 NGST, the expectations are that many will be spirals undergoing
 intense bursts of star formation (Ferguson and Babul
 \cite{fergusonbabul98}; Takamiya \cite{takamiya99}). To classify
two-armed spirals such as NGC 2997 (fig. 4) according to the
templates in figs. 5 and 7 is straightforward.
   According to the robustly determined Fourier spectra, its
 dust penetrated classification would be H2$\beta$. The point
of this investigation is that even if a large
 percentage of these high z systems are optically as irregular as
\object{NGC
 922}, only two low order modes (m=1 and m=2) are invariably required to
 describe the dust penetrated disk, in agreement with the ubiquity
 of low-order stellar modes found in our local Universe (Block and
 Wainscoat \cite{blockw91}; Block et al. \cite{blocketal94b},
 \cite{blocketal00}).

 \section{Conclusion}

 The utility of an optical imaging capability for NGST is currently
 being debated (a list of potential instruments is listed on the NASA
 Webpage www701.gsfc.nasa.gov/isim/science.htm). The performance of such
 an instrument can, to a limited degree, be extrapolated from the current
 capability of HST. Owing to the paucity of work in this area, it
 has been less clear what one hopes to learn from morphological
 studies undertaken in the $L$ and $M$ near-infrared bands that are
 the focus of much of the NGST science case.

 Our goal in this paper has been to argue (from a single
 illustrative example) that important morphological work will be
 achieved at these longer wavelengths. We have identified
 \object{NGC 922} as a reasonably representative ``archetypal''
 candidate similar in appearance to high redshift morphologically
 peculiar counterparts (see fig. 10).
 By simulating dust-penetrated
images of
 this object at z $\sim$ 1, and undertaking a detailed analysis of
 this synthetic data, we have shown that quantitative morphological
 investigation of such systems is well within the capabilities of
 the current NGST design.

 Perhaps more importantly, we have argued for the necessity of such
 observations in order to interpret the nature of optically
 peculiar systems at high redshifts. Optical observations of the
 high redshift Universe may not effectively trace underlying
 baryonic mass distributions or `galactic backbones', but rather
 focus for example, on the degree of symmetry/asymmetry of the
 dusty Population I component. This is especially true in systems
 undergoing large bursts of star formation, where the focus in the
 optical or UV is on OB associations and HII regions. A crucial
 lesson from our local Universe is that galaxies on opposite ends
of the Hubble tuning fork may
 present almost identical {\it mass}
 distributions,  and belong to the same dust-penetrated class. This result
 is clearly seen in figs. 11 and 12. It is not, a
 priori, possible to predict what galactic backbones might look
 like when the disks are dust penetrated at low, and at high, redshift.  At
 redshifts greater than 1.5, substantial mass could be locked into
 `old' (ages greater than one or two Gyr) stellar populations at
 surface brightness levels that are too faint to be visible on deep
 HST images (Abraham \cite{abraham98}). Near-IR imaging with NGST
 may well provide the most effective means for probing both evolved
 and dust-obscured stellar components in such systems.




 \begin{acknowledgements}

 It is a pleasure to thank the referee, Dr A. Moorwood, as well as Garth
Illingworth and Stephane Charlot for
 some very helpful comments.  AS is a Visiting Astronomer at the Infrared
 Telescope Facility,
             which is operated by the University of Hawaii under
             contract from the National Aeronautics and Space
             Administration.
 DLB, IP and AS are indebted to the Anglo-American Chairman's Fund
 Educational Trust. A note of deep appreciation is expressed to Mrs
 M. Keeton and the Board of Trustees. Dr Rolf Chini is also warmly
thanked for his input. This research is partially
 supported by the Mexican Foundation CONACYT under the grant No.
 28507-E.

 \end{acknowledgements}


\newpage

 \begin{figure}
           \special{
              hscale=60 vscale=60
              hoffset=-60 voffset=-95}
 \caption{A NASA IRTF near-infrared (K$'$)
 mosaic of NGC 922. An inner bisymmetric spiral, with wide open arms,
 lies
 within a prominent arc. The near-infrared disk of this optically
 chaotic specimen is remarkably regular: in fact, the inner spiral
 betrays the existence of arm
 modulation --  a diagnostic feature of several grand design galaxies such
as
 Messier 81.} \label{fig1} \end{figure}

 \begin{figure}
           \special{
               hscale=40 vscale=40
              hoffset=5 voffset=-35}
 \caption{850 $\mu m$ SCUBA contours overlayed on the K$'$ image. The
 contours start at 2$\sigma$ and increase in steps of 2$\sigma$.
 The peak lies at 27.5 mJy/beam (beam=$14\farcs5$ FWHM) and
 the noise level about 1.7 mJy/beam ($\sim$ 14$\sigma$).
 The
 integrated flux is 36 mJy, corresponding to a dust mass of $\sim$
 10$^{7}$ M$_{\odot}$.} \label{fig2}

 \end{figure}

 \begin{figure}
           \special{
              hscale=45 vscale=45
              hoffset=-15 voffset=-50}
 \caption{\object{NGC 922}. Upper left: \object{NGC 922}, which
 optically falls outside the Hubble Classification scheme, imaged
 at K$'$ with the NASA IRTF at Mauna Kea. Only two low-order modes
 (m=1 and m=2) are required to quantify the structure in the
 stellar disk (see upper right, where the m=1 and m=2 contours,
 determined from the inverse Fourier transform, are overlayed on
 the groundbased K$'$ image). Middle left: A simulated one hour
 image of \object{NGC 922} with an 8m NGST at a redshift z=0.7 (L
 band). Middle right: The L band image, with m=1 and m=2 contours
 overlayed. Bottom right and left show the galaxy redshifted to
 z=1.2 (M band) with and without contours. These M-band postage stamp
FITS images are only 3$''$ on a side. For simulations
 illustrated here, a value of q$_0$=0.1 is assumed. Notice the
 power of the Fourier method to maintain the ubiquity of first and
 second modes in stellar disks (Block et al. \cite{blocketal94b}),
 even for \object{NGC 922}.}
 \label{fig3}
 \end{figure}

 \begin{figure}
           \special{
              hscale=45 vscale=45
              hoffset=-15 voffset=-50}
 \caption{\object{NGC 2997}. The original groundbased K$'$ image is
 seen at upper left. Upper right: Evensided m=2 contours determined
 from the inverse Fourier transform are overlayed on the groundbased
 K$'$ image. Middle left: A simulated image of \object{NGC 2997} with
 an 8m-NGST at redshift z=0.7 (L-band). Middle right: The L band
 image, with contours (determined from the inverse Fourier transform)
 overlayed. Bottom right and left show the galaxy redshifted to
z=1.2 (M
 band) with and without contours. The M-band images are only $\sim$
2.5$''$ on a side.  Note the remarkable preservation of
 pitch angle with increasing redshift, and the power of the Fourier
 method to delineate spiral arms at restframe K$'$ even at z=1.2.}
 \label{fig4}
 \end{figure}

 \begin{figure}
           \special{
              hscale=62 vscale=62
              hoffset=-55 voffset=-15}
 \caption{ Old disks may be grouped into three principal archetypes: the
tightly wrapped $\alpha$ class, an intermediate $\beta$ class and an open
$\gamma$ bin. These dust penetrated classes are related to the rate of shear
in the stellar disk, as determined by $A/\Omega$, where $A$ is the first
Oort constant and $\Omega$ is the circular frequency. Actual deprojected
K$'$ images of $\sim$ 50 H1 and H2 galaxies have been superimposed to
generate this template. Galaxies of class H1 (the first Fourier harmonic)
present one dominant arm with a pitch angle belonging to one of the three
classes $\alpha$, $\beta$ or $\gamma$. Galaxies of class H2 (the second
Fourier harmonic) present two prominent spiral arms of pitch angle class
$\alpha$,
$\beta$ or $\gamma$.}
 \label{fig5}
 \end{figure}

 \begin{figure}
           \special{
              hscale=45 vscale=45
              hoffset=-15 voffset=-50}
 \caption{In our local Universe, one of the most well studied
lopsided spirals is
NGC 1637, seen imaged here in the near-infrared. The pitch angle of the
lopsided arm is $\sim$35$^{\circ}$; the arm belongs to the
wide open $\gamma$ class. In NGC 922, the single arc as well as the
inner bisymmetric spiral
in fig. 1 all belong to the $\gamma$ bin.}
 \label{fig6}
 \end{figure}

 \begin{figure}
           \special{
              hscale=40 vscale=40
              hoffset=-5 voffset=-3}
 \caption{ In this dust penetrated
   z$\sim 0$ tuning fork template,
 galaxies are binned according to three quantitative
   criteria: H$m$, where $m$ is the dominant Fourier harmonic
   (illustrated here are the two-armed H2 family); the pitch angle
   families $\alpha$, $\beta$ and $\gamma$ determined from the Fourier
   spectra, and finally the bar strength, derived from the
   gravitational potential (and not ellipticity) of the bar. Early type
   `b' galaxies \object{NGC 3992}, \object{NGC 2543}, \object{NGC 7083},
   \object{NGC 5371} and \object{NGC 1365} are
   distributed within all three families ($\alpha$, $\beta$ and $\gamma$).
  Adapted from Buta and Block (\cite{butablock01}).}
 \label{fig7}
 \end{figure}

 \begin{figure}
           \special{
              hscale=73 vscale=73
              hoffset=15 voffset=-65}
 \caption{The Fourier spectra of the K$'$ image of \object{NGC 922}
 at its unredshifted distance is shown at top.  The middle row
 shows the  Fourier spectra when \object{NGC 922} is moved to
 redshifts z=0.7 (L band) and z=1.2 (M band), respectively. We
 assume that a class 8m NGST has been used for securing the images,
 and that the on source integration time is one hour. We adopt a
 deceleration parameter of q$_0$=0.1. The bottom row shows the
 Fourier spectra for the same redshift values, but assuming a
 different cosmology with q$_0$=0.5. A remarkable similarity in the
 restframe K$'$ images of \object{NGC 922} is found, independent of
 redshift and of the deceleration parameters assumed.
Note that the key signature for spiral arm modulation of the m=2 components
$p>0$ and $p<0$ is beautifully retained into higher redshift space. }
\label{fig8}

 \end{figure}

 \begin{figure}
           \special{
              hscale=73 vscale=73
              hoffset=15 voffset=-65}
 \caption{As in fig. 8. Illustrated in the uppermost two
 panels are restframe K$'$ Fourier spectra of the grand design
 Sc spiral galaxy \object{NGC 2997}.
 The middle row
 shows the  Fourier spectra when \object{NGC 2997} is moved to
 redshifts z=0.7 (L band) and z=1.2 (M band), respectively.
 We adopt a
 deceleration parameter of q$_0$=0.1. The bottom row shows the
 Fourier spectra for the same redshift values, but assuming a
 different cosmology with q$_0$=0.5.}
 \label{fig9}

 \end{figure}

 \begin{figure}
           \special{
              hscale=45 vscale=45
              hoffset=-15 voffset=-50}
 \caption{\object{NGC 922} could well serve as a local Rosetta stone
for morphologically peculiar systems in our higher redshift Universe.
Apparent chaos reigns supreme in this deep optical B-band image of NGC
922;
dust-penetrated imaging, however, shows that a simple two-armed spiral
(betraying the signature of arm modulation found in several grand design
prototypes such as Messier 81) is largely responsible
for the dynamics in the stellar backbone of NGC 922 whose Fourier
spectra are presented in fig. 8.}
 \label{fig10}
 \end{figure}

\begin{figure}
           \special{
              hscale=65 vscale=65
              hoffset=15 voffset=-65}
 \caption{Lessons from our local Universe show that the dust
           penetrated Fourier spectra of galaxies on opposite sides on
           the Hubble tuning fork may be almost identical. Seen here
           are the Fourier spectra generated from a K$'$ image of NGC
           718 (Hubble type a). Compare these spectra to those of NGC 309
           (Hubble type c) in fig. 12.} \label{fig11} \end{figure}

\begin{figure}
           \special{
              hscale=65 vscale=65
              hoffset=15 voffset=-65}
 \caption{The Fourier spectra generated from a type c galaxy (NGC 309)
           at restframe K$'$ are remarkably similar to those presented
           in fig. 11 of NGC 718 (type a). Both NGC 309 and NGC 718
            belong to the same dust
penetrated bin $\beta$, and near-infrared images of these galaxies may
           be found in fig. 7.  The galactic backbones of NGC 718
           and NGC 309 are very similar,
           despite the fact that optically, these two galaxies are
           located on opposite ends of the Hubble tuning fork.} \label{fig12} \end{figure}

 \end{document}